\documentclass[11pt,a4paper]{article}
\usepackage{moriond,epsfig}
\usepackage[usenames]{color}

\def\Journal#1#2#3#4{{#1} {\bf #2}, #3 (#4)}

\def\ApJ{\em ApJ.}
\def\MNRAS{\em MNRAS}

\def\be{\begin{equation}}
\def\ee{\end{equation}}
\def\bea{\begin{eqnarray}}
\def\eea{\end{eqnarray}}

\begin{document}
\vspace*{4cm}
\title{The Distribution of Dark Matter in the Universe on scales 
of 10$^{10}$ M$_\odot$ to 10$^{15}$ M$_\odot$~\footnote{Talk presented 
by L.~Teodoro at the XXXIXth Rencontres de Moriond on 
{\em Exploring the Universe}, La Thuile, Italy, March 28 - April 4, 2004.}}

\author{ Lu\'{\a i}s Teodoro and Michael~S.~Warren~\footnote{teodoro@lanl.gov(LT),
msw@lanl.gov(MSW)}}

\address{Theoretical Division, Los Alamos National Laboratory\\
Los Alamos, New Mexico, USA}

\maketitle\abstracts{ The use of parallel computers and increasingly
sophisticated software has allowed us to perform a large suite of
N-body simulations using from $10^8$\,to $10^9$ particles.  We will
report on our recent convergence tests of the halo mass function,
N-point correlation functions, power spectrum and pairwise velocity
from very large high resolution treecode N-body simulations. Rather
than basing results on just one or two large simulations, now one can
investigate the role of different numerical and physical effects on
the statistics used to characterize the mass distribution of the
Universe. }

\section{Introduction}
Over the last two decades cosmological $N$-body simulations have
played a crucial role in the study of the formation and evolution of
cosmic structure. In this approach, the initial statistical properties
of the fluctuations, at some early stage of the universe, are set
using linear theory. $N$-body simulations are then used to evolve the
structure into the deeply nonlinear regime.  This dynamical state is
then compared with the large-scale structure in the galaxy data-sets.

The impressive progress achieved in the observational front with the
completion of very large surveys such as 2dF and SDSS poses a clear
challenge to the numerical work in cosmology: the precision of the
predictions provided by the current $N$-body experiments have to be of
the order of a few percent.

\section{Numerical Simulations}
The cosmological model adopted is a low-density, flat, $\Lambda$CDM
Universe with the parameters $\Omega_m=0.3$, $\Omega_\Lambda = 0.7$
and $h=0.7$.  The power spectrum of the initial conditions was set up
using the output transfer function of CMBFAST \cite{sz}, assuming
$\Omega_b= 0.04$ and a normalisation of $\sigma_8=0.9$. The entire
suite of simulations were performed using the Hashed Oct-Tree code
(HOT)~\cite{ws}, a parallel tree code with periodic boundary
conditions.  The simulation parameters are listed in Table 1.

\begin{table}[htbp]
\begin{center}
\caption{Suite of simulations}
\vspace{0.2in}
\begin{small}
\begin{tabular}{lccccccl}
\hline \hline
\\
    &            &                  &     & \#     & L$_{box}$ &  $\epsilon$  & ~~~~\,m$_{p}$\\
Run &$\Omega_m$  & $\Omega_\lambda$ & $h$ & Part. & [Mpc $h^{-1}$] & [kpc $h^{-1}$] & ~~~[M$_{\odot}$] \\
\\
\hline
\\
{\color{Red}{dtd10}}        & {\color{Red}{0.30}} & {\color{Red}{0.70}} & {\color{Red}{0.70}} & 
{\color{Red}{512$^3$}} & {\color{Red}{1536}} & {\color{Red}{98.0}} & {\color{Red}{2.3 $\times$ 10$^{11}$}} \\
{\color{Green}{dtd11}}      & {\color{Green}{0.30}} & {\color{Green}{0.70}} & {\color{Green}{0.70}} & 
{\color{Green}{512$^3$}} & {\color{Green}{768}}  & {\color{Green}{49.0}} & {\color{Green}{2.8 $\times$ 10$^{10}$}}\\
{\color{Blue}{dtd12}}       & {\color{Blue}{0.30}} & {\color{Blue}{0.70}} & {\color{Blue}{0.70}} & 
{\color{Blue}{512$^3$}} & {\color{Blue}{384}}  & {\color{Blue}{24.5}} & {\color{Blue}{3.5 $\times$ 10$^{9}$}}\\
{\color{YellowOrange}{ej1}} & {\color{YellowOrange}{0.30}} & {\color{YellowOrange}{0.70}} & {\color{YellowOrange}{0.70}} & 
{\color{YellowOrange}{512$^3$}} & {\color{YellowOrange}{192}} & {\color{YellowOrange}{12.3}} & {\color{YellowOrange}{4.4 $\times$ 10$^9$}}\\
{\color{RedViolet}{ef10}}   & {\color{RedViolet}{0.30}} & {\color{RedViolet}{0.70}} & {\color{RedViolet}{0.70}} & 
{\color{RedViolet}{512$^3$}} &  {\color{RedViolet}{96}}  &  {\color{RedViolet}{4.9}} & {\color{RedViolet}{5.5 $\times$ 10$^8$}} \\
\\
{\color{cyan}{ef4}}         & {\color{cyan}{0.30}} & {\color{cyan}{0.70}} & {\color{cyan}{0.70}} & {\color{cyan}{768$^3$}} & 
{\color{cyan}{1152}} & {\color{cyan}{65.5}} & {\color{cyan}{2.8 $\times$ 10$^{11}$}}\\
{\color{magenta}{ef7}}      & {\color{magenta}{0.30}} & {\color{magenta}{0.70}} & {\color{magenta}{0.70}} & 
{\color{magenta}{768$^3$}} & {\color{magenta}{288}}  & {\color{magenta}{14.0}} & {\color{magenta}{4.4 $\times$ 10$^9$}}\\
\\
{\color{NavyBlue}{ei6}} & {\color{NavyBlue}{0.30}} & {\color{NavyBlue}{0.70}}& {\color{NavyBlue}{0.70}} & 
{\color{NavyBlue}{1024$^3$}}& {\color{NavyBlue}{768}}  & {\color{NavyBlue}{24.5}} & {\color{NavyBlue}{3.5 $\times$ 10$^{9}$ }}\\
\\ 
\hline \hline
\end{tabular}
\end{small}
\end{center}
\label{tab:nbody-historical}
\end{table}

\section{Analysis}

\begin{figure}
  \centering
  \vskip-0.01in\relax
  \includegraphics[width=3.0in]{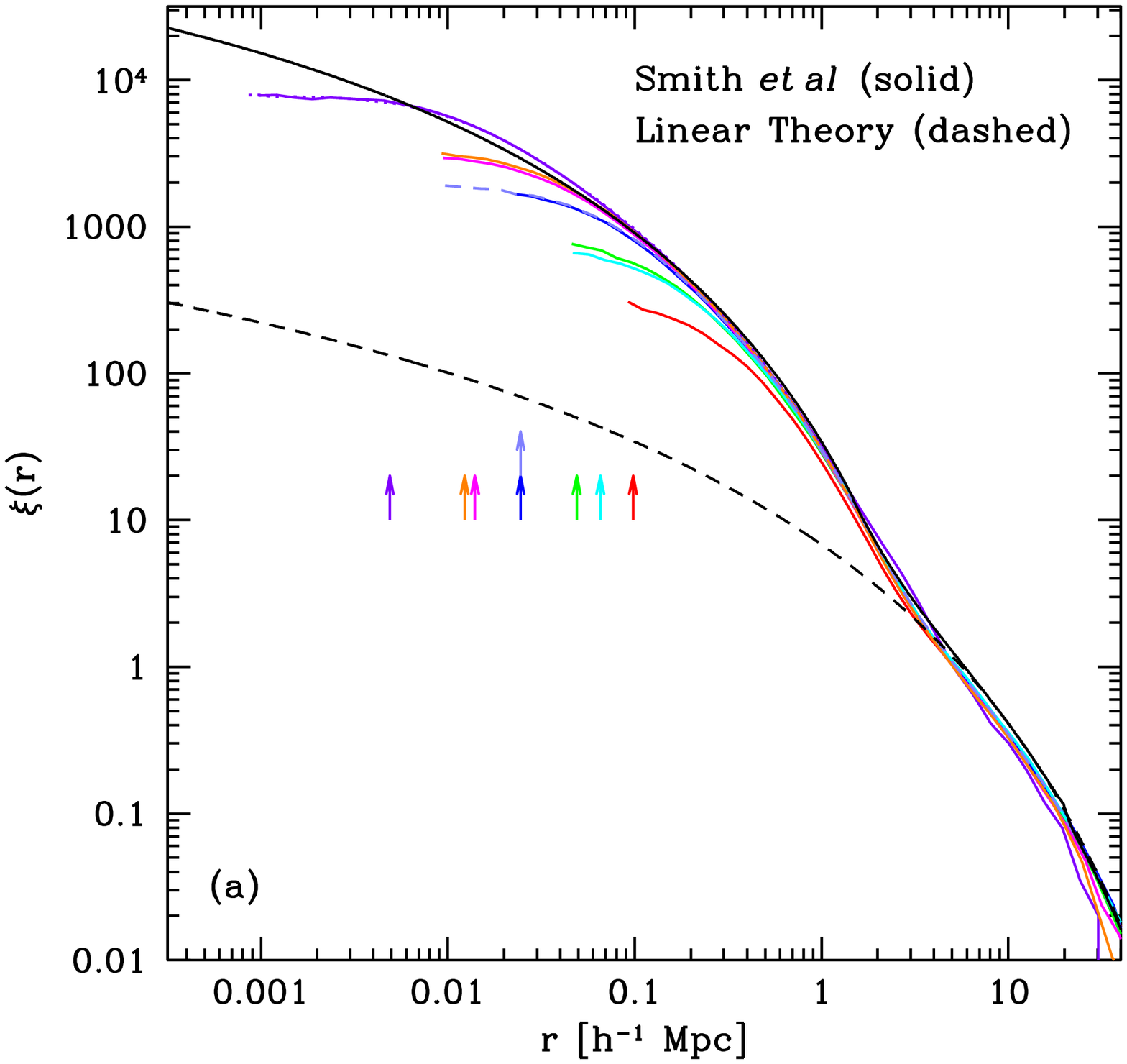}
  \includegraphics[width=3.0in]{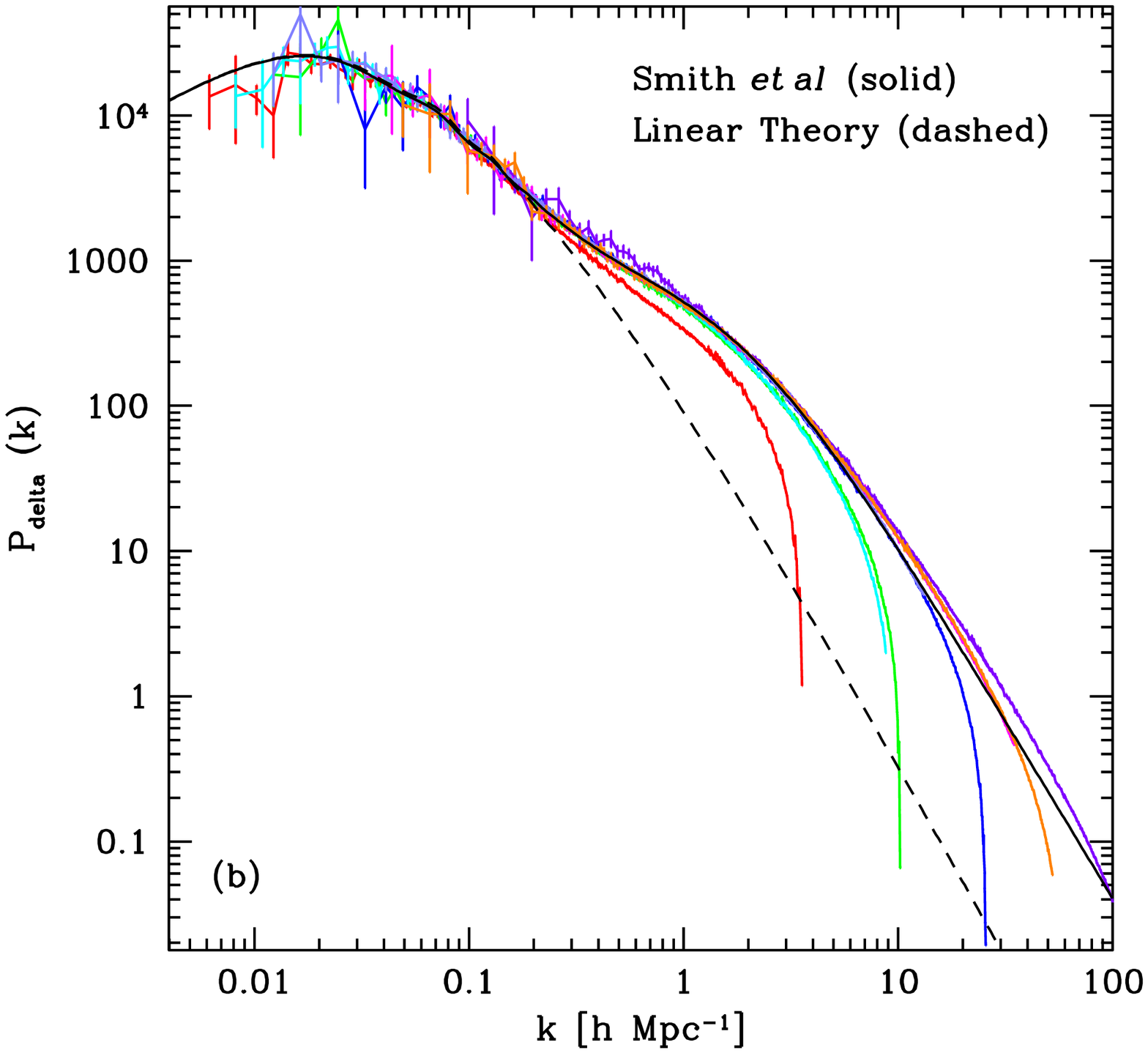} 
  \includegraphics[width=3.0in]{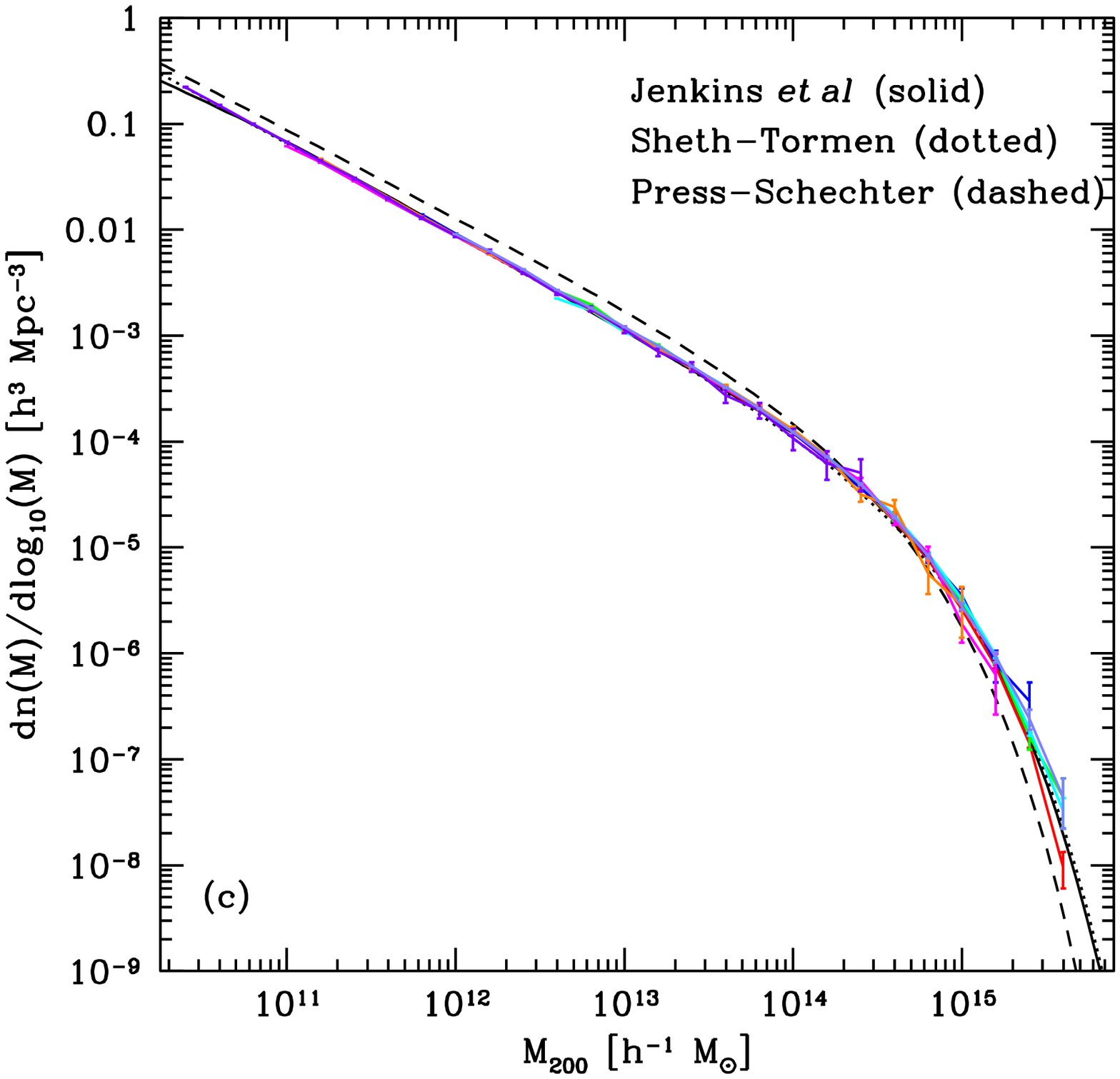} 
  \includegraphics[width=3.0in]{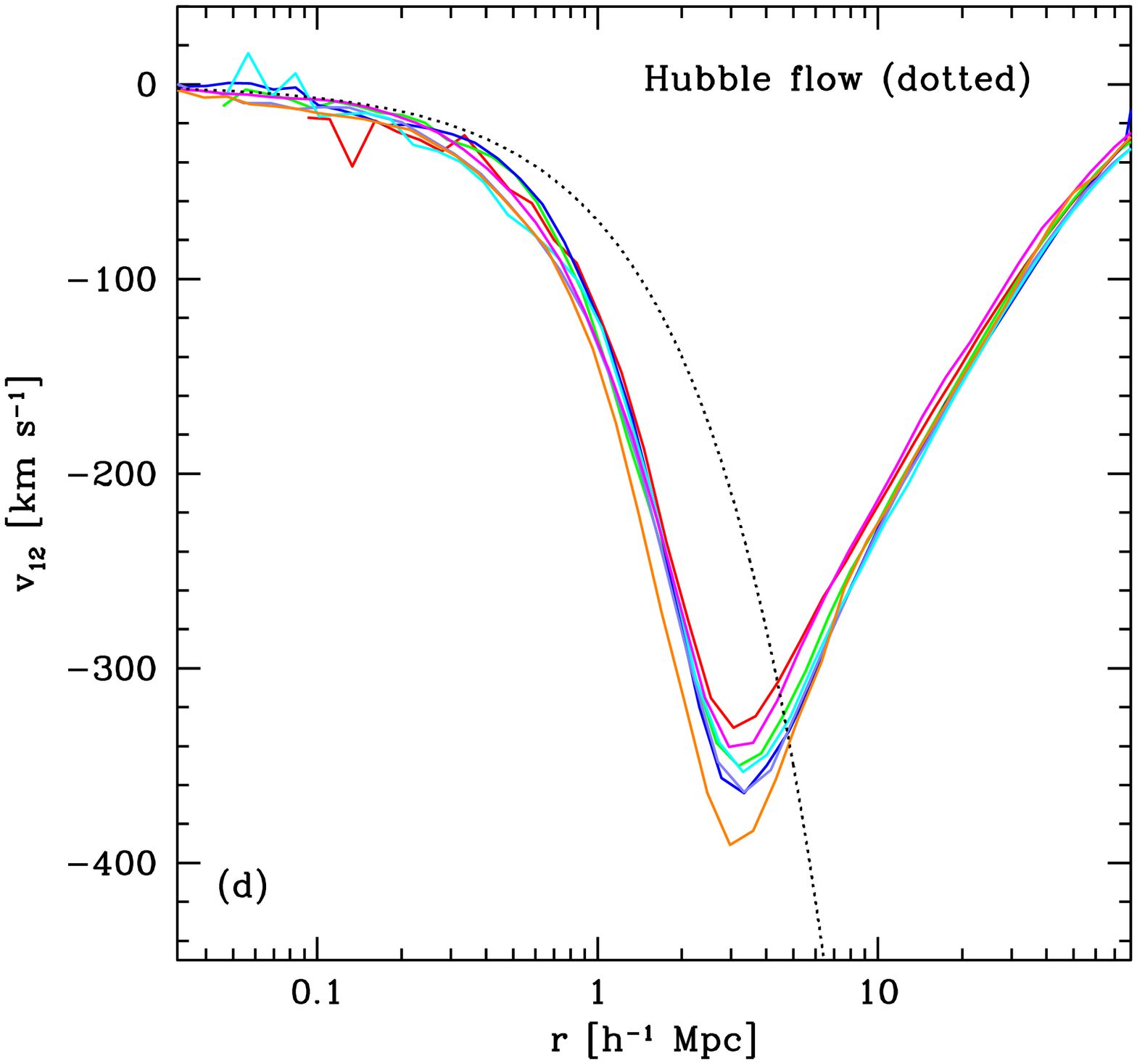}
  \caption{(a) Two point correlation function, (b) power spectrum, (c)
mass function and (d) mean pairwise velocity. See color map in Table
1.  The arrows in the top left panel indicate the gravitational
softening. In the same panel, we also show the linear theory (dashed
line) and the Smith {\it et al} (solid line) two-point correlation
function fitting for the assumed cosmological model. The solid, dotted
and dashed lines in panel (c) show the Sheth-Tormen, Jenkins {\it
et\,al} and Press-Schechter~$^{10}$~mass functions,
respectively. Finally, in panel (d) the dashed line represents the
Hubble flow.}
  \label{fig:all}
\end{figure}

The main purpose of this analysis is to understand how changes in mass
resolution, particle number and box size affect some well known
estimators applied in Large Scale Structure studies.  For each
cosmological volume we estimate the two-point correlation function,
the power spectra, the mass function and the mean pairwise velocity,
see Figure~\ref{fig:all}.  The color map is shown in Table 1.

The two-point correlation function is presented in
Figure~\ref{fig:all}\,(a). On small scales, the amplitude of the
two-point correlation function is suppressed by softening gravity and
lack of small scale power in the initial conditions. This affects the
estimates in a range which roughly extends up to five times the size
of the smoothing kernel [see arrows in top left panel of the
abovementioned figure].  The flattening of the two-point correlation
function seems to disappear once the mass resolution (softening)
increases (decreases). On scales where $\xi \approx 1 $ all the
simulations show a rather good agreement among themselves. The box
size only looks to matter on scales well within the linear regime
($\xi \ll 1$) where the smaller boxes present a slightly smaller
amplitude. The solid (dashed) lines represent the two-point Smith
{\it et al}\,~\cite{smith_etal} (linear) correlation function.

To measure the power spectra of our simulations we applied the
technique detailed in Jenkins {\it et al}\,~\cite{jenkins_etal1998} and
Smith {\it et al}\,~\cite{smith_etal}.  In doing so discreteness and
grid effects are taken in account and removed from the final
estimation. As expected, the power spectra of the different $N$-body
simulations show the same features as the two-point correlation
function. The analytical expression for the non-linear power spectra
due to Smith {\it et al}\,~\cite{smith_etal} and the linear power
spectra are shown by the solid and dashed black lines,
respectively. Futher details can be found in Warren \&
Teodoro~\cite{wt}.

We used the friends-of-friends algorithm~\cite{defw} to identify dark
matter halos. This halo finder depends on just one parameter, $b$,
which defines the linking length as $bn^{-1/3}$~where $n$ is the mean
particle density; we follow the prescription of Jenkins {\it et
al}\,~\cite{jenkins_etal2001} and set $b=0.2$. Figure~(c) demonstrates
that the agreement between the analytical expressions of Jenkins {\it
et al}~~\cite{jenkins_etal2001} (solid line) and Sheth and
Tormen~\cite{st1999} (dotted line) and our estimates of the mass
function is excellent. The small departures seen in the mass range $M
> 10^{14} h^{-1} M_\odot$ are mainly caused by small number
statistics.

Finally, Figure~\ref{fig:all} (d) shows the mean pairwise velocity,
$v_{12}(r)$, as function of pair separation. The Hubble line, given by
$ v_{Hubble} = -Hr$, is also shown.  The mean pairwise velocity
vanishes at the smallest separations resolved in our simulations. In
the subset of simulations which better probe small scales (small
volume and larger mass resolution) $v_{12}(r)$ closely follows the
Hubble line up to $\sim 300$~kpc $h^{-1}$. This indicates that
structures on such scales are close to relaxation.  The mean pairwise
velocity reaches a minimum in scales comparable to the correlation
length [see Figure 1(d)]. At larger scales the $v_{12}(r)$ intersects
the Hubble line and, at very large separations it decays to zero, in
accordance with the principle of large-scale homogeneity and isotropy.

Some simulations in this suite have been further analyze in Seljak \&
Warren~\cite{seljak2004}.

\section*{Acknowledgments}
We thank A.R.~Jenkins, G.~Evrard, K.~Abazajian and S.~Habib for help
and useful discussions.  This work was performed under the auspices of
the U.S. Dept.  of Energy, and supported by its contract
$\#$W-7405-ENG-36 to Los Alamos National Laboratory. Simulations were
performed on the Space Simulator Beowulf cluster at Los Alamos
National Laboratory.

\end{document}